\documentclass[aps,pra,reprint,twocolumn,superscriptaddress,showpacs,showkeys,a4paper]{revtex4-1}

\usepackage{amsmath,amssymb,amstext}
\usepackage[usenames,dvipsnames]{color}
\usepackage{graphicx,subfigure,relsize}
\usepackage{bm,bbold,braket}

\definecolor{mygreen}{rgb}{0,0.5,0}
\definecolor{myblue}{rgb}{0,0,0.75}
\definecolor{mymagenta}{cmyk}{0,1,0,0.12}

\newcommand{\eq}[1]{\begin{equation} #1 \end{equation}}
\newcommand{\eqa}[1]{\begin{eqnarray} #1 \end{eqnarray}}

\newcommand{\br}{\boldsymbol{r}}
\newcommand{\bk}{{\bm k}}
\newcommand{\be}{\boldsymbol{e}}
\newcommand{\bsig}{\boldsymbol{\sigma}}
\newcommand{\bh}{\boldsymbol{h}}

\newcommand{\hh}{\hat{h}}

\newcommand{\ud}{\mathrm{d}}
\newcommand{\ue}{\mathrm{e}}
\newcommand{\Tr}{\mathrm{Tr}}

\newcommand{\vect}[1]{\boldsymbol{#1}}

\begin{document}

\title{Tomography of band insulators from quench dynamics}

\author{Philipp Hauke}
    \email{philipp.hauke@uibk.ac.at}    
        	\affiliation{Institut f\"ur Quantenoptik und Quanteninformation,\"Osterreichische Akademie der Wissenschaften, Technikerstr.~21A, A-6020 Innsbruck, Austria}
	\affiliation{	Institut f\"ur Theoretische Physik, Universit\"at Innsbruck, Technikerstr.~25, A-6020 Innsbruck, Austria}
\author{Maciej Lewenstein}
	\affiliation{ICFO -- Institut de Ci\`{e}ncies Fot\`{o}niques, Parc Mediterrani de la Tecnologia, E-08860 Castelldefels, Spain}
    	\affiliation{ICREA -- Instituci{\'o} Catalana de Recerca i Estudis Avan\c{c}ats, Lluis Companys 23, E-08010 Barcelona, Spain}
\author{Andr\'e Eckardt}
    \affiliation{Max-Planck-Institut f\"ur Physik komplexer Systeme, N\"othnitzer Str.\ 38, D-01187 Dresden, Germany}

\date{\today}

\begin{abstract}

We propose a simple scheme for tomography of band-insulating states in one- and two-dimensional optical lattices with two sublattice states. In particular, the scheme maps out the Berry curvature in the entire Brillouin zone and extracts topological 
invariants such as the Chern number. The measurement  relies on observing---via time-of-flight 
imaging---the time evolution of the momentum distribution following a sudden quench in the 
band structure. We consider two examples of experimental relevance: the Harper model with
$\pi$-flux and the Haldane model on a honeycomb lattice. Moreover, we illustrate the performance of 
the scheme in the presence of a parabolic trap, noise, and finite measurement resolution.  
\end{abstract}



\maketitle

Band insulators are a fascinating form of quantum matter. Their very existence relies on band 
structure and quantum statistics and the geometric phase inherent in their wave function gives rise 
to intriguing physics: 
Topological quantum numbers, such as the Chern index or the $Z_2$ invariant, separate 
them into classes with fundamentally different behavior 
\cite{Kitaev2008,Ryu2010}, some of which support chiral or helical edge modes \cite{Hasan2010,Qi2011}.Only recently, first evidence of such topological insulators has been found in solid-state materials 
\cite{Kim2012Topol,Zhang2013,Gehring2013} and photonic systems \cite{Rechtsman2012}. 
The recent creation of artificial gauge fields in optical lattices \cite{Aidelsburger2011,Jimenez2012,
Struck2012,Atala2012,Struck2013,Parker2013,Aidelsburger2013,Miyake2013} makes it seem likely that 
topological insulators will be realized in the near future also in highly tunable systems of 
ultracold atoms. However, the experimental characterization of apparently structureless band 
insulators poses a challenge. 
So far, there are schemes that are designed to measure specific topological properties of the system. 
For example,  Zak's phase, i.e., the Berry phase acquired during the adiabatic motion 
along a path through the Brillouin zone (BZ), was measured recently from Bloch oscillations
\cite{Atala2012}. Also, the location of Dirac cones was mapped out in a honeycomb lattice using Landau-Zener transitions 
\cite{Tarruell2012,Lim2012,Lim2013}. Other 
proposed schemes are designed to directly measure either the Chern number (from density profiles 
\cite{Umucalilar2008}, wave-packet dynamics \cite{Price2012,Atala2012,
Abanin2012,Dauphin2013,Liu2013Topol}, time-of-flight (TOF) imaging \cite{Zhao2011,Liu2013Topol,Burrello2013}, or 
unidirectional TOF imaging with single-site resolution \cite{Wang2013}), or to probe the presence of 
chiral edge modes (via transport measurements \cite{Goldman2010,Goldman2012c,Killi2012,Killi2012b}, in particular  using quench-based schemes, 
or Bragg-scattering \cite{Liu2010,Stanescu2010,Buchhold2012,Goldman2012}). A method allowing for a 
full tomography of a band insulator has so far only been proposed for a specific experimental 
realization of a topological insulator based on spin-dependent hexagonal 
lattices \cite{Alba2011}. 

Here, we propose a simple scheme for the complete tomography of band-insulting states in
one-dimensional (1D) and two-dimensional (2D) optical lattices that is not restricted to a specific 
system. In particular, the scheme allows for mapping-out the Berry curvature as a function of 
quasimomentum and for measuring the Chern number. Our scheme is based on the momentum-resolved 
monitoring---via TOF imaging---of the dynamics following an abrupt quench 
in the band structure. In the following, we  first introduce the basic protocol
underlying our method, and then discuss two relevant applications: the $\pi$-flux Harper model 
\cite{Harper1955,Hofstadter1976}, and the Haldane model \cite{Haldane1988}.

\begin{figure}
	\centering
\includegraphics[width=0.49\textwidth]{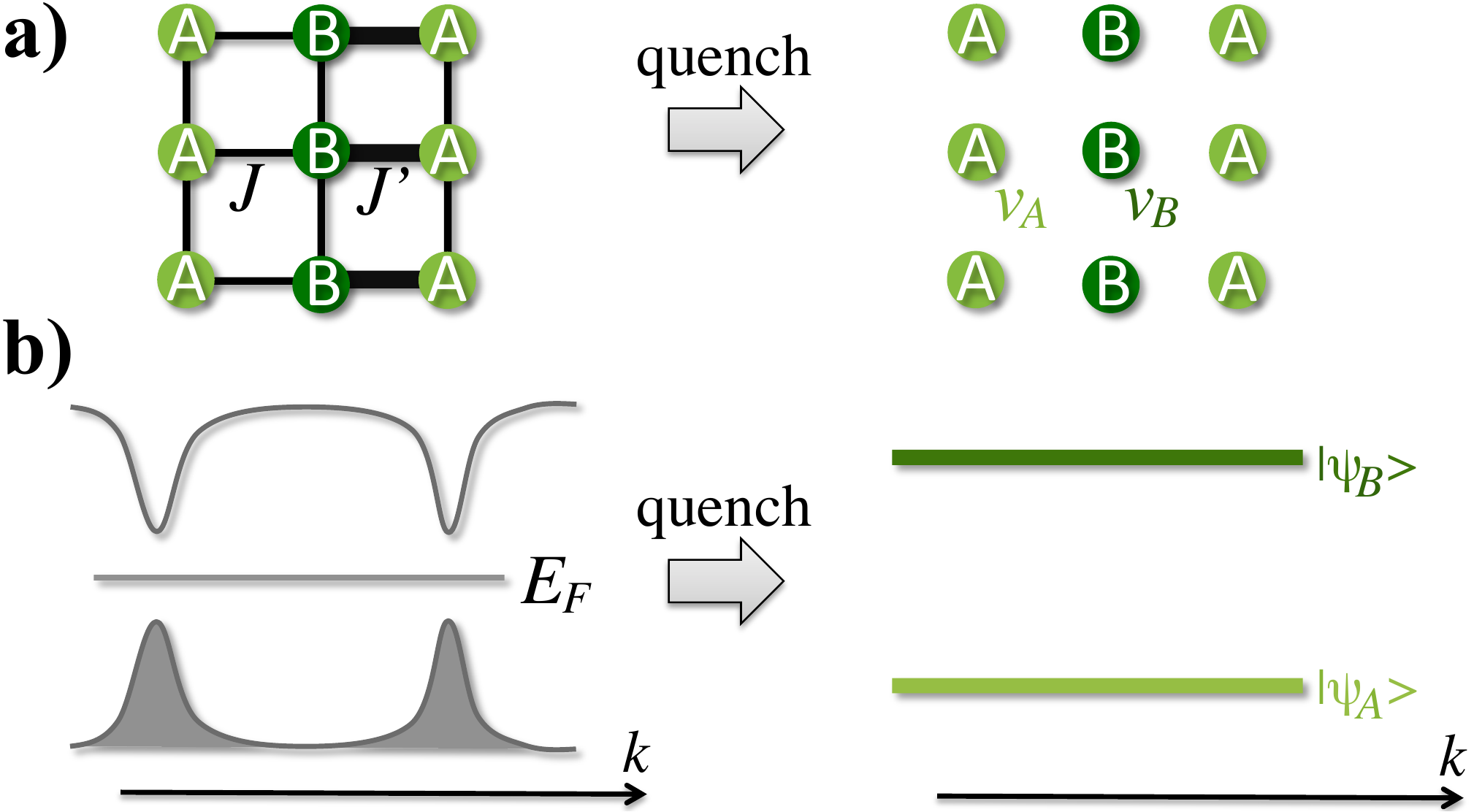}
\caption{
General procedure.
(a) A lattice with two sublattice states is quenched abruptly such that both sublattices are 
energetically separated by $\hbar\omega=v_A-v_B$ and tunneling is suppressed (drawn example: a 
square lattice with alternating tunneling matrix elements $J$ and $J^\prime$). 
(b) Initially, the system is a band insulator with the lower band occupied completely. 
At every quasimomentum $\bk$, the two-dimensional state space is represented by a Bloch sphere, with 
the state of the lower band lying at $-\hat{\bh}(\bk)$. 
With the quench, two flat bands are created, onto which the state is projected. The resulting dynamics corresponds to a rotation around the $z$-axis of the Bloch sphere with the same  frequency
$\omega$ for every $\bk$. Monitoring this dynamics in momentum space 
allows for reconstruction of the initial position on the Bloch sphere, giving a complete 
tomography of the initial band-insulating state.  
\label{fig:procedure}	}
\end{figure}

\emph{Scheme for the tomography of band insulators.---} Consider spin-polarized (i.e., non-interacting) fermions in a 2D optical lattice. In each elementary cell $\ell$, the lattice shall have two sublattice states $s=A,B$, located at $\br_{\ell s}$
[see Fig.~\ref{fig:procedure}(a), left]. The corresponding tight-binding Hamiltonian 
is characterized by matrix elements $h_{\ell's',\ell s}$ that obey the translational symmetry of 
the lattice. The diagonal terms refer to on-site potentials, $h_{\ell s,\ell s}\equiv v_s$, and the 
off-diagonal matrix elements describe tunneling between near neighbors. Thanks to the translational 
symmetry, the Hamiltonian is diagonal with respect to quasimomentum $\bk$. With respect to the basis 
states $|\bk s\rangle \propto \sum_{\ell}\ue^{-i\bk\br_{\ell s}}|\ell s\rangle$ it is represented by
a $\bk$-dependent $2\times2$ matrix 
$h_{s's}({\bm k})=\sum_{\ell,\ell'} h_{\ell' s',\ell s} \ue^{-i (\br_{\ell's'}-\br_{\ell s})\bk}$, 
which we decompose as $h_{s's}(\bk)\equiv h_0(\bk)\delta_{s's}+\bh(\bk)\bsig_{s's}$. Here,
$\bsig_{s's}$ denotes the vector of Pauli matrices in sublattice space. For  every quasimomentum $\bk$,
the 2D sublattice space defines a Bloch sphere, with north and south
pole given by $|\bk A\rangle$ and $|\bk B\rangle$, respectively. The two eigenstates
$|\bk \pm \rangle$ lie at $\pm\hat{\bh}(\bk)$ on this Bloch sphere, where 
$\hat{\bh}(\bk)\equiv \bh(\bk)/|\bh(\bk)|\equiv \big(\sin(\vartheta_{\bk})\cos(\varphi_{\bk}),
\sin(\vartheta_{\bk})\sin(\varphi_{\bk}),\cos(\vartheta_{\bk})\big)$. Their energies
$\varepsilon_\pm(\bk)=h_0(\bk)\pm |\bh(\bk)|$ define the band structure of the lattice.

We consider the system to be in a band-insulating state with
complete occupation of the single-particle states of the lower band, $|\bk-\rangle=
\sin(\vartheta_\bk/2)|\bk A\rangle-\cos(\vartheta_\bk/2)\ue^{i\varphi_\bk}|\bk B\rangle$ \cite{footnoteUnipartite}. 
Here, we assume that the gap is much larger than the temperature, allowing to neglect thermal excitations, which would decrease the observed contrast. 
This band-insulating state is represented by the map $\hat{\bh}(\bk)$ from the first BZ onto the Bloch sphere. The topological
 properties of this map determine the properties 
of the system. Our aim is to design a feasible measurement scheme that allows  for a reconstruction of $\hat{\bh}(\bk)$. 

The momentum (not quasimomentum) distribution of the band insulator, which is obtained from 
TOF images taken after suddenly switching off the lattice potential, is given  by $n(\bk) =
f(\bk)|\langle{ \bk-}|\bk A\rangle+\langle {\bk-}|\bk B\rangle|^2 
=f(\bk)[1-\sin(\vartheta_\bk)\cos(\varphi_\bk)]$. Here, $f(\bk)$ is a 
broad envelope function given by the momentum distribution of the Wannier function; the 
expression in square brackets possesses the periodicity of the reciprocal lattice. Unfortunately, 
$n(\bk)$ does not provide sufficient information to reconstruct $\hat{\bh}(\bk)$ or, equivalently,
both $\vartheta_\bk$ and $\varphi_\bk$. In order to obtain the missing information, at the 
measurement time $t_m$ the system shall be subjected to an abrupt quench
$h_{\ell's',\ell s}\to h'_{\ell's',\ell s}$, such that  a potential off-set
$v'_A-v'_B\equiv\hbar\omega$  between $A$ and $B$ sites is created and tunneling suppressed
[see Fig.~\ref{fig:procedure}(a), right]. 
In quasimomentum representation, the Hamiltonian is now characterized by a constant vector
$\bh'(\bk)\approx (\hbar\omega/2) \be_z$ generating a rotation around the $z$-axis of the Bloch 
sphere with frequency $\omega$. 
Starting from the band-insulating state, this dynamics is, thus, captured simply by replacing
$\varphi_\bk\to\varphi_\bk+\omega (t-t_m)$. This leads to an observable dynamics in the momentum distribution 
\begin{equation}
\label{eq:nkt}
n(\bk,t)=f(\bk)[1-\sin(\vartheta_\bk)\cos(\varphi_\bk+\omega(t-t_m))],
\end{equation}
whose oscillatory time dependence directly reveals both $\varphi_\bk$ and
$\sin(\vartheta_\bk)=1-|\hat{h}_z(\bk)|^2$. The time-dependence of $n(\bk,t)$ allows us to 
reconstruct $\hat{h}_x(\bk)$, $\hat{h}_y(\bk)$, as well as $|\hat{h}_z(\bk)|$ from the amplitude and 
the phase of the oscillations. It is sufficient to consider data for  $\bk$ 
from the first BZ;  the Wannier envelope $f(\bk)$ does not 
spoil the measurement as it just gives an irrelevant overall prefactor for each value of $\bk$. For 
a full tomography, it remains to reveal the sign of $\hat{h}_z(\bk)$. Since the overall sign is not 
important, one has to determine those lines where $\hat{h}_z(\bk)$ changes sign. These lines can be 
clearly identified by a characteristic cusp-like behavior of $|\hat{h}_z(\bk)|$,
$|\hat{h}_z(\bk)|\propto |\bk-\bk_\text{sign-change}|$, which sharply contrasts with the smooth 
variation of $\hat{h}_z(\bk)$  as it results from
tunneling between near neighbors. 

Moreover, $h_z(\bk)$, including its sign, can also be 
measured via band mapping: After abruptly switching on a strong potential off-set lifting $B$
with respect to $A$ sites as before, the lattice is switched off without waiting time at a 
slow rate such that quasimomentum is mapped onto momentum. Absorption images after TOF 
reveal then a momentum distribution where the $A$ ($B$) population, corresponding to the lowest
(first excited) band, is mapped onto the first (second) BZ.

\emph{Edge states in the Harper model with $\pi$-flux.---} Once $\hat{\bh}(\bk)$ is reconstructed,
one can infer whether the system supports edge 
modes or not, by invoking the bulk-boundary correspondence. According to the procedure derived in
Ref.~\cite{Ryu2002}, for that purpose one has to identify a closed path $\bk(\lambda)$ in
$\bk$-space such that $\hat{\bh}(\bk(\lambda))$ lies in a plane $\mathcal{E}$ that contains the 
origin (i.e., it lies on a great circle of the Bloch sphere). If the unit vector $\hat{\bh}$ describes a closed 
circle around the origin when moving along the path, the system does possess zero-energy edge modes; 
if not, it does not. 

Motivated by recent experiments \cite{Aidelsburger2013,Miyake2013}, let us consider the example of a 
square lattice with nearest-neighbor tunneling and with a flux of $\pi$ (half a flux quantum) per 
plaquette, which can be generated via laser-assisted tunneling or lattice shaking \cite{Goldman2013}.
We consider a gauge where the tunneling matrix elements in $y$-direction alternate between $-J$ 
and $+J$ when moving through the lattice in $x$-direction, giving two inequivalent sublattices
$s=A,B$. Additionally, we assume different on-site energies $v_A=\Delta/2$ and $v_B=-\Delta/2$, and that the tunneling matrix element in $x$-direction alternates between $-J'$ and 
$-J$ [Fig.~\ref{fig:procedure}(a)]; both can be achieved by a superlattice in $x$-direction. The 
extent of the first BZ is given by $\pi$ ($2\pi$) in $k_x$ ($k_y$) direction. The quasimomentum-space Hamiltonian of this model is characterized by 
\eq{
\label{eq:HarperModelHalfFlux}
\vect{h}({\vect{k}}) = (-J - J^\prime \cos(2 k_x), J^\prime \sin(2 k_x), -2 J \cos(k_y) + \Delta)\,.
}
The two parameters $\Delta/J$ and $J^\prime/J$ allow us to explore various situations with qualitatively different band structures. 
Most notably, for $J=J^\prime$ and $\Delta<2J$, one finds two Dirac cones  lying at
$k_x=\pm\pi/2$ and $k_y=\arccos(\Delta/2J)$. 
For imbalanced tunneling matrix elements $J' \neq J$, a band gap opens at the Dirac points and 
edge states appear if $J'>J$.  
We focus here on the case $\Delta=0$ (for $\Delta\neq0$ see~\cite{suppl}). 
In this case, by choosing $k_y^0=\pi/2$ ($\hh_z=0$), we can confine $\hat{\bh}_{k_y^0}(k_x)$ to a plane that contains the origin of $\hat\bh$'s Bloch sphere, a necessary condition for observing edge modes at zero energy \cite{Ryu2002}. 
Edge states (in the equivalent system with open boundary conditions in $x$-direction) do appear if the origin is encircled by $\hat{\bh}_{k_y^0}(k_x)$ \cite{Ryu2002}.

\begin{figure}
	\centering
\includegraphics[width=0.5\textwidth]{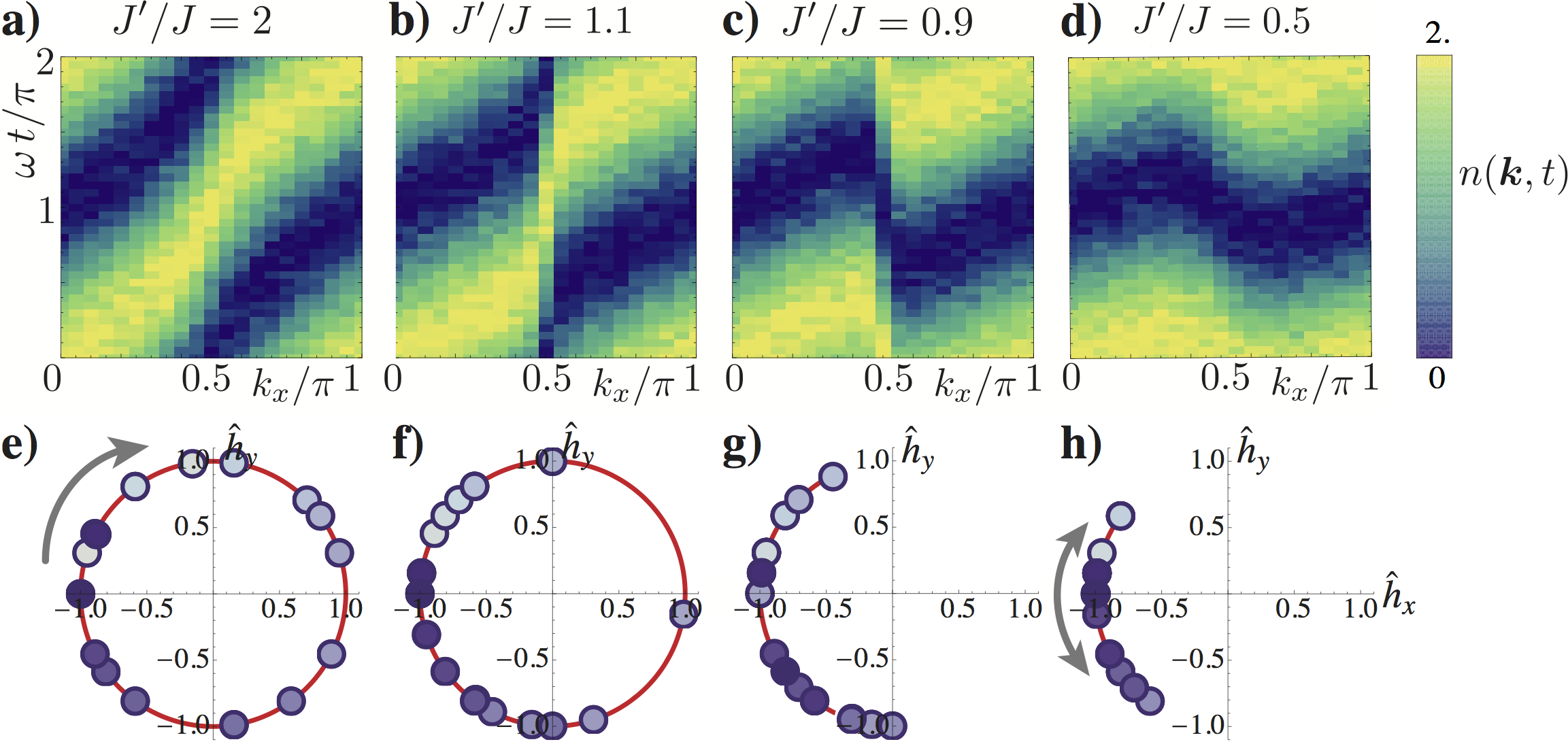}	
\caption{
(a-d) Time-resolved TOF images for the $\pi$-flux Harper model ($\Delta=0$ and $k_y^0=\pi/2$). To 
mimic a realistic experiment, 
we added normal-distributed noise with a standard deviation of 0.1 of the average signal and assumed 
a limited resolution of 21 points along $k_x$ and 41 time points. 
(e-f) If edge states are supported, the temporal maximum as a function of quasimomentum winds around 
the time period once (panels a,b), and $\hat{\vect{h}}_{k_y^0}(k_x)=(\hat{h}_x,\hat{h}_y,0)$ 
describes a unit circle around the origin (e,f), contrary to when no edge state is supported (c,d 
and g,h, respectively). 
Red line: exact case. Bullets: values extracted from the noisy, resolution-limited data of the upper row, using Eq.~\eqref{eq:nkt} [$k_x$ from $0$ (light) to $\pi$ (dark)].
 \label{fig:nkoftimeHarperTimeEvolution}
	}
\end{figure}
In an experiment, we wish to reconstruct $\hat{\bh}_{k_y^0}(k_x)$ from the dynamics following a sublattice quench in order to conclude whether the system possesses edge states or not.
A typical result will look like the upper row of Fig.~\ref{fig:nkoftimeHarperTimeEvolution}, where 
we plot $n({\bk},t)$ at $k_y^0=\pi/2$, $\Delta=0$, and four values of $J'/J$. In order to illustrate 
the robustness of our method, we have contaminated $n({\bk},t)$ with normal-distributed uncorrelated 
noise, with a standard deviation of ten percent of the average signal. 
Furthermore, we assumed a mediocre experimental resolution of 21 points in the first BZ along $k_x$ and 41 points in time.
We can reconstruct $\hat{\vect{h}}_{k_y^0}(k_x)$ from ${n}(k_x,k_y^0;t)$ 
simply by identifying $\varphi({\bk})$ with the position of the first maximum in time and
$\sin(\vartheta_\bk)=1-|\hat{h}_z(\bk)|^2$ with the difference between maximum and minimum. 
The resulting graphs $\hat{\vect{h}}_{k_y^0}(k_x)=(\hat{h}_x,\hat{h}_y,0)$, 
plotted in the lower row of Fig.~\ref{fig:nkoftimeHarperTimeEvolution}, clearly reveal the presence 
or absence of edge states: Edge states, expected for the two plots on the left where $J'>J$, are 
clearly indicated by data points describing a circle around the origin. 
Qualitatively, one can see this information already in the time evolution of $n({k_x,k_y^0};t)$ in 
the upper row: If the band insulator supports edge states, 
the maximum winds around the time period. 

Our scheme also permits to monitor the topological transition of the model happening when $\Delta$ 
exceeds $2J$, where for $J'=J$ both Dirac cones merge (see~\cite{suppl}, where  edge modes for open
boundary conditions are discussed also).

\emph{Measuring Berry curvature and Chern number in a Haldane-like system.---} Edge currents are topologically protected only if the associated integer Chern number, given by the 
integral of the Berry curvature over the whole BZ, is finite. In the above example, this is not the 
case, since the edge modes always appear in counter-propagating pairs located at the two
Dirac cones. We now turn to a lattice model where a finite Chern number 
can be found and demonstrate how our scheme can be used to map out the Berry curvature in 
quasimomentum.

The Berry curvature of the lower band is given by 
\eq{
\label{eq:Berrycurvature}
\Omega_{-}(\bk)=\frac{1}{2}\left(\partial_{k_x}\hat\bh\times\partial_{k_y}\hat\bh\right)\cdot\hat\bh 
\;,}
and is readily obtained from $\hat{\bh}(\bk)$. It describes the polarizability \cite{Xiao2010} 
and the anomalous Hall conductivity \cite{Haldane2004} also at lower filling.  
The Chern number reads
\eq{
\label{eq:nC}
w_-=\frac{1}{2\pi}\int\ud^2k\,\Omega_{-}(\bk)\;.
}
It counts how often $\hat{\bh}(\bk)$ wraps around the Bloch sphere when $\bk$ covers the full 
first BZ. It is proportional to the Hall conductivity of the completely filled band and indicates
the presence of robust chiral edge modes \cite{Hasan2010}.

\begin{figure}
	\centering
\includegraphics[width=0.5\textwidth]{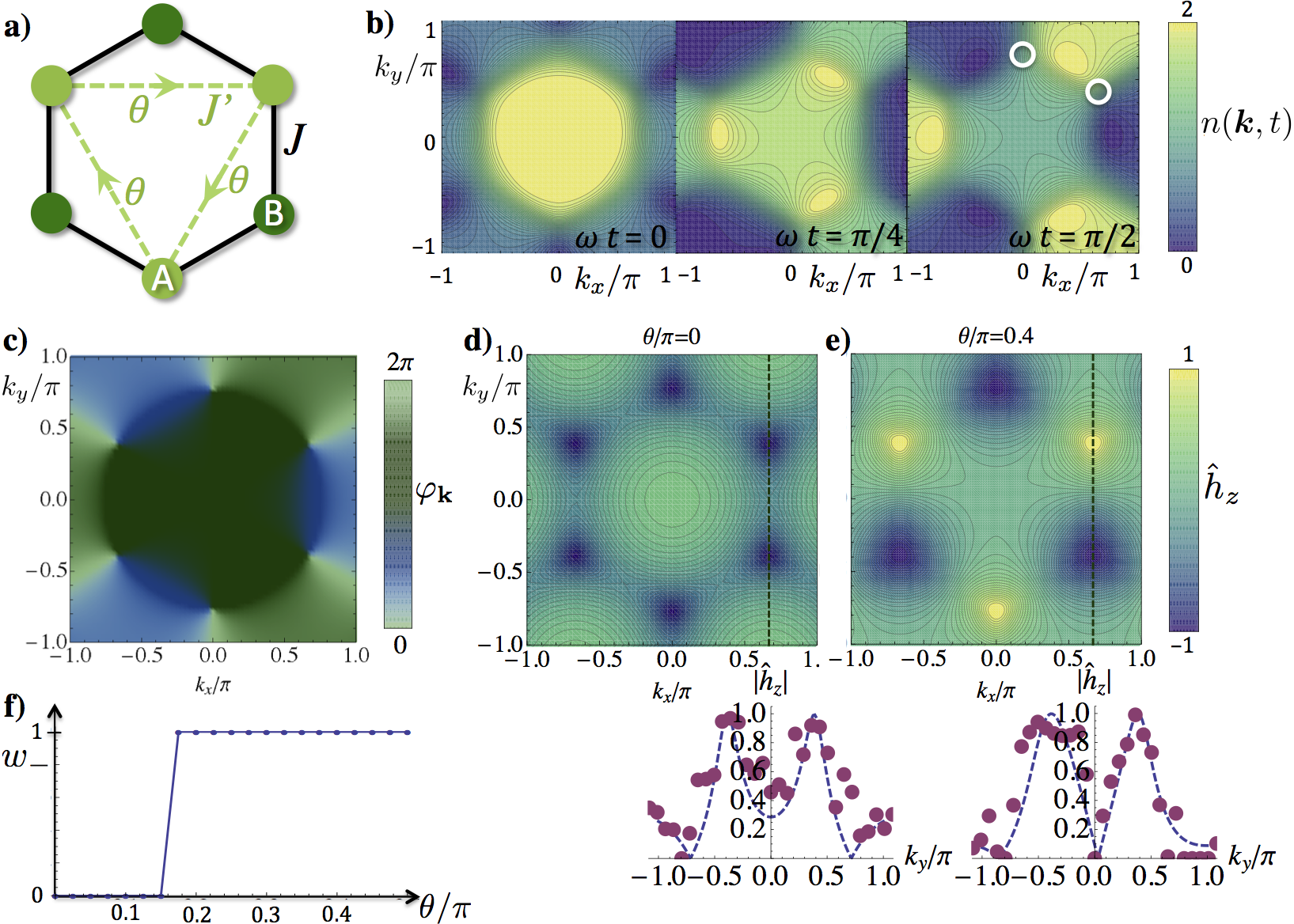}
\caption{
{(a)} Haldane-like lattice model with real tunneling parameter $J$ and complex tunneling parameter
$J^\prime\ue^{i\theta}$. 
{(b)} $n({\bk},t)$ changes strongly with time ($J^\prime=0.3 J$, $\theta=0.4\pi$). 
{(c)} The winding of the phase $\varphi_{\bk}$ is opposite around the two Dirac points (white circles in b), independently of $\theta$.   
{(d,e)} In the topologically trivial phase (d, $J'=0.3J$ and $\theta=0$), the plotted quantity
$\hh_z$ has the same sign at the two Dirac cones, while in the topological phase (d, $J'=0.3J$ and
$\theta=0.4\pi$) it has opposite sign; this sign change is clearly visible as a kink in the measured quantity $|\hh_z|$ (lower plots showing $|\hh_z|$ along the dashed lines; lines: ideal case, bullets: data for trapped system with normal-distributed noise of variance 0.05, 
using realistic parameters \cite{Aidelsburger2013,Struck2013}, $J/\hbar=2\pi\, 0.26\,\rm{kHz}$, $\omega=2\pi\, 10\,\rm{kHz}$, trapping frequency $2\pi\, 50\,\rm{Hz}$, lattice spacing $380\,\rm{nm}$).  
{(f)} The coarse-grained Chern number computed following Ref.~\cite{Fukui2005} reproduces the exact result already for a limited resolution of $4\times 4$ reciprocal lattice points.
\label{fig:Haldane} }
\end{figure}

Let us consider the Haldane-like model sketched in
Fig.~\ref{fig:Haldane}(a) with the lower band completely filled. The atoms live on a 
honeycomb lattice \cite{SoltanPanahi2011,Tarruell2012} with real tunneling matrix elements $J$ 
between nearest neighbors (NNs) and complex tunneling matrix elements $J^\prime\ue^{i \theta}$ with 
Peierls phase $\theta$ between next-nearest neighbours (NNNs). The model can be realized, e.g., in a shaken optical 
lattice~\cite{Hauke2012c}. For $J'=0$, the vector $\bh(\bk)$ lies in the $xy$-plane and the band 
structure possesses two Dirac points where $\bh(\bk)=0$. For finite NNN tunneling, $J'>0$, $\hat{h}_z(\bk)$
acquires a finite value and a direct band gap opens at the Dirac cones (though one still has an 
indirect band touching). While for $J'=0$ the unit vector $\hat{\bh}(\bk)$ was confined to the equator, when approaching the Dirac points it will now visit either the north or the 
south pole, depending on the sign of $\hat{h}_z(\bk)$. 
The unit vector $\hat{\bh}(\bk)$ can only wrap around the Bloch sphere, as required for a finite 
Chern number, if $\hat{h}_z(\bk)$ has opposite sign at the two Dirac points such that it visits both 
poles. We find that, while for $|\theta|<\theta_c$ the system
is a trivial band insulator, it becomes a topological Chern insulator, characterized by a Chern
number $|w_-|=1$, once $|\theta|>\theta_c\approx0.18\pi$.

As exemplified in Fig.~\ref{fig:Haldane}(b) for an initial state in the topological phase
 ($\theta=\pi/2$), $n({\bk},t)$ changes its pattern strongly as a function of time. From this 
dynamics we can
extract the position of $\hat\bh(\bk)$ on the Bloch sphere [Fig.~\ref{fig:Haldane}(c-e)]. The sign 
change of $\hh_z$ for $\theta/\pi>0.18$ [Fig.~\ref{fig:Haldane}(e)] between the two Dirac cones 
identifies a finite Chern number $|w_-|=1$. Although within our scheme one can directly 
measure only the absolute value $|\hh_z|$, the sign change in $\hh_z(\bk)$ can clearly be identified 
from the pronounced kink where $|\hh_z|$ touches zero, either between the two Dirac cones, 
indicating the topological phase [Fig.~\ref{fig:Haldane}(e)], or elsewhere, as in the trivial 
phase [Fig.~\ref{fig:Haldane}(d)]. The fact that the sign change of $h_z(\bk)$ can occur 
between the Dirac cones, where the band gap is largest, allows to indirectly identify a topological 
band structure even if the system is not in a perfect band insulating state. Namely, thermal 
excitations or small deviations from unit filling are relevant mainly near the Dirac cones, 
and not where the sign change occurs. 

In a realistic situation, the resolution of $n({\bk},t)$ will be restricted. 
Approximating the Chern number~\eqref{eq:nC} by a sum over differences is unreliable close to the topological transition (see~
\cite{suppl}). Much better results are obtained by the gauge-invariant description in terms of effective field strengths developed by Fukui \emph{et al.} \cite{Fukui2005}. 
In the Supplemental Material~\cite{suppl}, we show how their formula can be expressed in terms of $\hat\bh$. 
Since this method enforces an integer result, it gives the exact answer already for very small numbers of reciprocal lattice points. 
We demonstrate this in Fig.~\ref{fig:Haldane}(f), where we use for $n(\bk)$ only $4\times 4$ coarse-grained pixels in the first BZ and take only 10 time steps. 
We again obtain $\sin\vartheta_{\bk}$ from the maximal amplitude of the data points and $\varphi_{\bk}$ from the position of the maximum (for such a low resolution, the sign of $h_z$ can be obtained from band mapping).  
Remarkably, even for this extremely resolution-limited situation, the Chern number can be reproduced accurately. 

A natural question concerns the role of the trapping potential. As shown in~\cite{suppl}, a harmonic trap modifies the measured momentum distribution \eqref{eq:nkt} roughly by a prefactor
$\left(\frac{\mu_0-\epsilon_-(\bk)}{\mu_0-\epsilon_{\min}}\right)^2 {\rm sinc}^2 \left(\frac{(\mu_0-\epsilon_-(\bk))t}{2}\right)$, with Fermi energy $\mu_0$ and band minimum $\epsilon_{\min}$.
The first term describes the reduced contrast of modes with high energy, since these are only populated in the central region of the trap, 
and the second term captures dephasing during the post-quench time-evolution, caused by the 
spatially varying potential energy. 
For realistic parameters, both effects are small, and the proposed scheme works reliable even in the presence of a trap~\cite{suppl} (cf.\ the lower panels of Fig.~\ref{fig:Haldane}).
A newer generation of experiments may enable avoiding a spatially varying trapping potential altogether \cite{Gaunt2013}. 

{\emph{Discussion, conclusion, and outlook.---} 
The robust and simple method for the tomography of band insulators described here is not restricted 
to the two discussed  examples. It can be applied to any 1D or 2D band-insulator 
with two states per elementary lattice cell---interesting examples include the Su--Schrieffer--Heeger
\cite{Su1979} or Rice--Mele model \cite{Rice1982}, which was recently realized in 
an optical lattice \cite{Atala2012}. Here, an interesting application would be to measure  a topological charge pump \cite{Wang2013b} to extract 
the Chern number quantizing the transport of matter. Moreover, the method can also be employed to 
measure systems with only partially-filled lowest band, and it provides a means to validate 
Hamiltonians synthesized for the purpose of quantum simulation. 
As an outlook, it will be interesting to generalize the scheme to lattices with more than two 
sublattice states and to include internal atomic states.

{\emph{Acknowledgements.---}} We acknowledge discussions with Alexander Szameit and Leticia Tarruell. 
This work was supported by the EU IP SIQS, EU STREP EQuaM, ERC AdG OSYRIS, ERC synergy grant UQUAM, Fundaci\'o Cellex, and SFB FoQuS (FWF Project No. F4006-N16).

\renewcommand{\theequation}{S\arabic{equation}}
\setcounter{equation}{0}
\renewcommand{\thefigure}{S\arabic{figure}}
\setcounter{figure}{0}

\vspace*{1cm}
{\begin{center}{\LARGE{\bf Supplemental material}}\end{center}}

In this Supplemental Material, 
(i)    we provide additional information on how the topological phase transitions in the Harper and Haldane models may be observed, 
(ii)   we adapt the method of Ref.~\cite{Fukui2005} to compute Chern numbers directly from the Hamiltonian in quasimomentum representation, using only a small number of grid points, and 
(iii)  we present analytical and numerical calculations including a possible harmonic trapping potential, showing that its presence is not detrimental to the proposed scheme.

\subsection{\label{sec:topologicaltr}\label{sec:spectrumHarper} Possible observation of topological phase transitions in the $\pi$-flux Harper model}

\begin{figure*}
	\centering
\includegraphics[width=\textwidth]{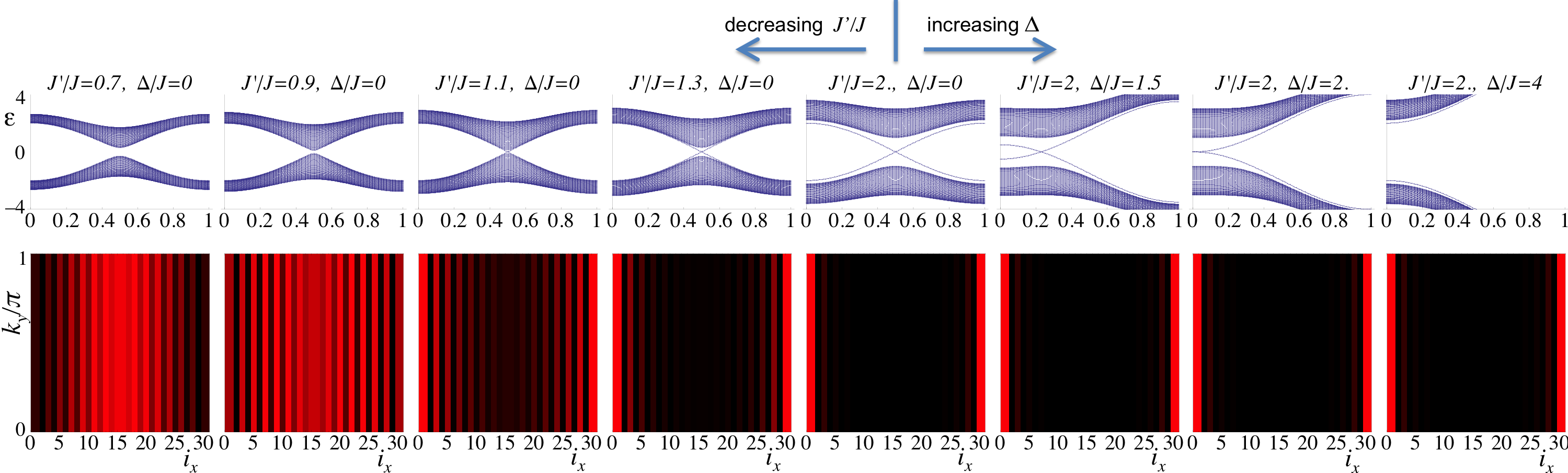}
\caption{
	  { Edge states in the $\pi$-flux Harper model,} for a cylindric geometry (open boundaries in $x$-direction with even number of sites, periodic in $y$-direction). 
	  {Upper row: Spectrum.} 
	  {Lower row: Occupation probability as a function of $k_y$.} Shown is the combined probability for the two modes closest to $\epsilon=0$ (for increased visibility, the color map is rescaled for each figure). 
	  Edge states appear when $J^{\prime}/J>1$. 
	  At $\Delta=2J$, the two points of minimal direct gap (the Dirac cones if $J^\prime=J$) merge, and as a result possible edge modes do not lie at zero energy for any $\Delta>2J$. 
	  \label{fig:edgeStatesHarperModel}
	  }
\end{figure*}
Here, we discuss the possibility of observing the topological phase transitions supported by the Harper model at half a flux quantum [as given by Eq.~(2) of the main text].
We first describe under which circumstances zero-energy edge states are to be expected. 
Afterwards, we turn to the question how these can be detected using the quench-based band tomography. 

The spectrum of the Harper model with $\pi$ flux is exemplified in Fig.~\ref{fig:edgeStatesHarperModel}, upper row, for several values of
$J^\prime/J$ and $\Delta/J$ (we consider an infinite lattice in $y$-direction, and $L=80$ sites in $x$-direction). 
In these spectra, edge modes can clearly be distinguished as lines that are separated from the bands. 
The lower row of Fig.~\ref{fig:edgeStatesHarperModel} shows the joint occupation of the two modes that come closest to zero energy (for $L=30$ for 
better visibility). These are the two edge modes at the opposite ends of the sample, if any edge states exist. 
In the absence of edge modes, the plotted modes are the bulk modes at the band boundary. 

As these figures show, at any $J^\prime \neq J$, a band gap opens. Edge states appear on both 
boundaries when $J^\prime>J$, i.e., when the edge is connected to the bulk by a weak link. 
Associated to each Dirac cone and boundary, there is exactly one edge mode with finite $k_y$ (note 
that only half of the first BZ, $k_y\in [0,\pi]$, is shown). The edge modes belonging to the two 
Dirac cones propagate in opposite directions, leading to a vanishing Chern number. 
When $\Delta>2J$, the Dirac cones merge at $k_y=0$, and as a result the edge modes do not cross the 
band gap. They approach the bulk energy bands with increasing $\Delta$.

\begin{figure}
	\centering
	\includegraphics[width=0.49\textwidth]{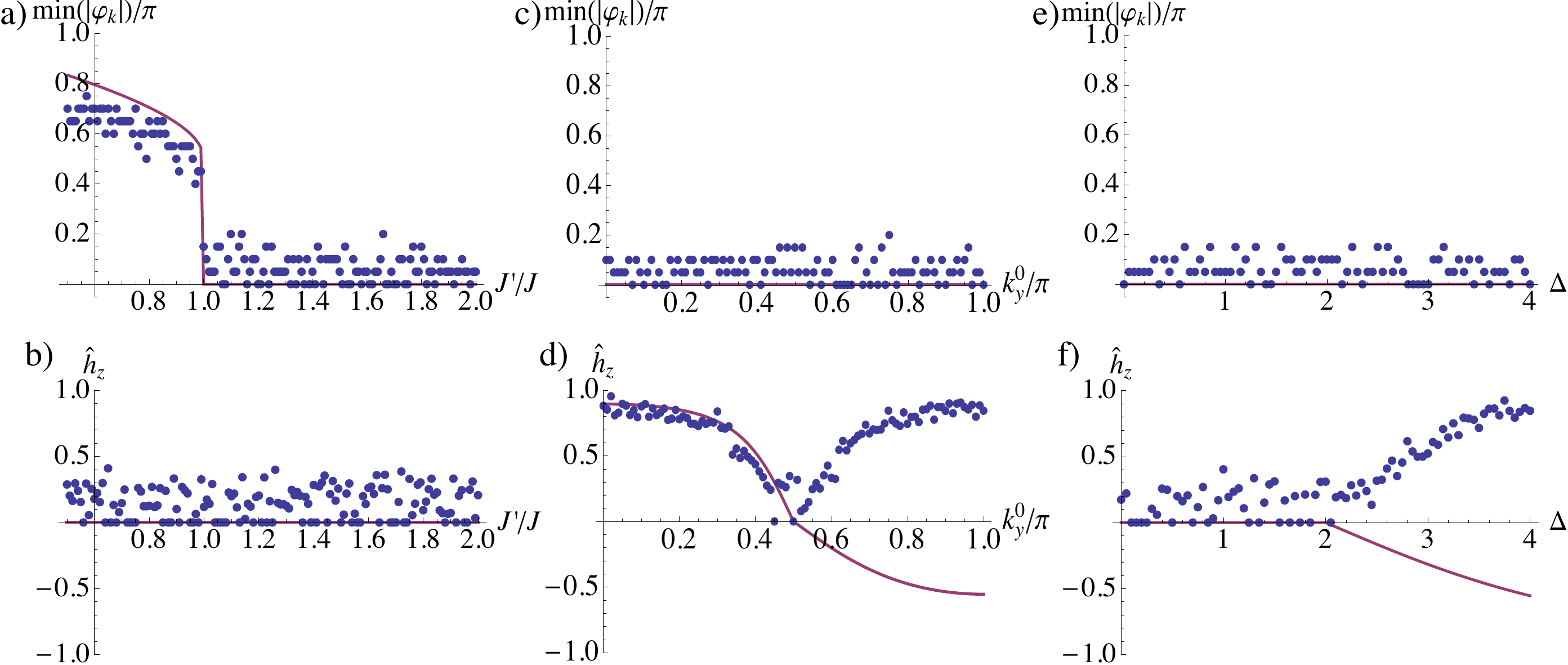}	
\caption{
	  {Topological phase transitions in the  $\pi$-flux Harper model.} 
  	{Upper row (a,c,e):}  
		Edge modes at $k_y^0$ are supported if $\hat{\vect{h}}_{k_y^0}(k_x)$ encloses a circle around the origin.
		Whether the circle is closed can be quantified by $\min|\varphi_{k_x,k_y^0}|$, even for noisy `data' similar to Fig.~2 of the main text. 
		{Bottom row (b,d,f):} The edge states are only at zero-energy if $\hat{h}_z=0$ (i.e., the parametric curve $\hat{\vect{h}}_{k_y^0}(k_x)$ encloses $\hat{\vect{h}}=0$). 
		The solid line represents the ideal prediction and the dots simulate a noisy experiment (which measures only the absolute value of $\hat{h}_z$). 
		{(a,b)} As a function of $J^\prime/J$, at $\Delta=0$ and $k_y^0=\pi/2$: zero-energy edge modes appear for $J^\prime/J>1$. 
		{(c,d)} As a function of $k_y^0$, at $J^\prime/J=2$ and $\Delta=0$: although edge modes may be supported at any $k_y^0$, their energy vanishes only at the position of the Dirac cone. 
		{(e,f)} As a function of $\Delta$, at $J^\prime/J=2$ and $k_y^0$ at the position of one of the Dirac points: when $\Delta>2J$, no zero-energy edge modes are supported.  
	  \label{fig:phaseTransitionHarperModel}}
\end{figure}

We now turn to the question how these topological transitions can be observed using the proposed scheme. 
We proceed analogous to Fig.~2 of the main text. First, we use Eq.~(2) to compute $n(\bk,t)$ on a finite grid in time and quasimomentum space within the first BZ.
To mimic a realistic experimental situation, we again add noise with standard deviation of 10\% of the average value. 
We use Eq.~(1) to extract from this data the characteristics of $\hat\bh(\bk)$ along a path in $\bk$-space with fixed $k_y^0$. 
The results are summarized in Fig.~\ref{fig:phaseTransitionHarperModel}. 

According to Ref.~\cite{Ryu2002}, the system supports zero-energy edge modes if the path $\hat\bh_{k_y^0}(k_x)$ is (i) closed and (ii) encircles $\hat\bh(\bk)=0$. 
In the considered model, the second condition can only be fulfilled if $\hat{h}_z(k_x,k_y^0)=0$, i.e., if $\Delta<2J$ and $k_y^0=\arccos(\Delta/2J)$. 
Otherwise, the system may still support edge modes, but these then do not lie at zero energy. 
The fulfilment of the first condition, the closing of the curve, can be identified by a small value of $\min|\varphi_{k_x,k_y^0}|$ [see Figs.~2(e-h) of the main text for examples]. 

Indeed, in the entire topological phase at $J'/J>1$, we find $\min|\varphi_{k_x,k_y^0}|\approx 0$ [shown in Fig.~\ref{fig:phaseTransitionHarperModel}(a), where $\Delta=0$ and $k_y^0=\pi/2$]. 
For these parameters, we moreover find $|\hat{h}_z|\approx 0$, indicating that the edge states lie at zero energy [Fig.~\ref{fig:phaseTransitionHarperModel}(b)]. 

As Fig.~\ref{fig:phaseTransitionHarperModel}(c) exemplifies for the point $\Delta=0$ and $J'/J=2$, edge modes actually exist for all $k_y^0$ as long as $J^\prime>J$ and $\Delta<2J$ (see Fig.~\ref{fig:edgeStatesHarperModel}). 
However, these modes lie at zero energy only at the Dirac point, which can be seen by a sharp kink of $|\hat{h}_z|$ as a function of $k_y^0$ [Fig.~\ref{fig:phaseTransitionHarperModel}(d)]. 
As illustrated in Fig.~\ref{fig:phaseTransitionHarperModel}(e,f) for $J'/J=2$, when increasing $\Delta$ to values larger than $2J$, we find $|\hh_z|>0$: the Dirac cones have merged and, although edge states still exist, they are no longer at zero energy.

\subsection{\label{sec:nCalaFukui}\label{sec:nCfromdifferencesums}Chern numbers in a Haldane-like model, extracted from discrete $k$-space grids}

In a realistic TOF image of $\bh(\bk)$, the resolution will be restricted. 
If data is available only on a discrete mesh $\{\bk_{m}\}$ of the BZ, one would typically approximate the integral in Eq.~(4) of the main text by a discrete sum, and the derivatives occurring in the integrand by a discrete difference. 
While this procedure gives reasonable approximations to the Chern number deep inside a given phase  (see Fig.~\ref{fig:nC_differencesums}), it is unreliable close to the topological phase transition (see also Ref.~\cite{Alba2011}). 
\begin{figure}
	\centering
\includegraphics[width=0.25\textwidth]{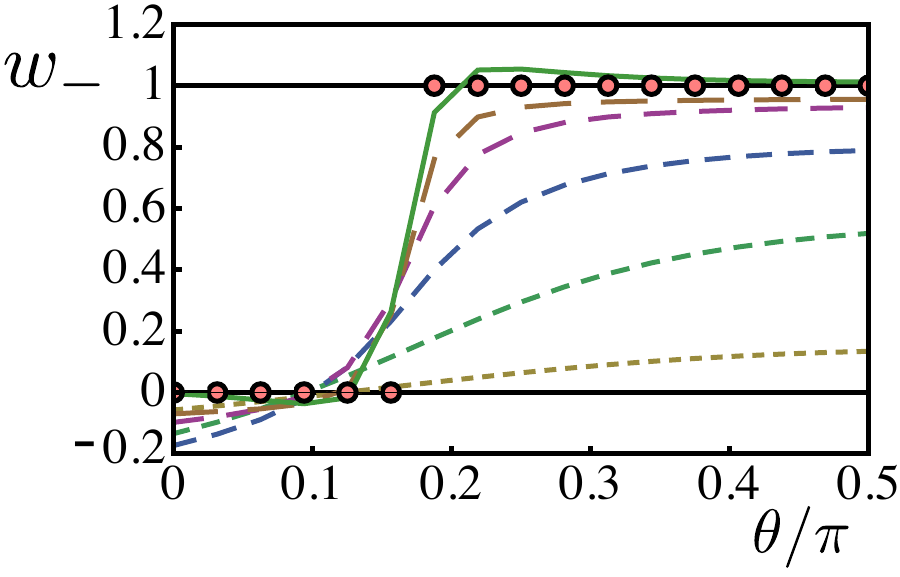}
\caption{
	  {Chern number of Haldane model [Fig.~3(a)], computed from difference sums over a $\bh(\bk)$ with finite resolution  ($J'=0.3J$).}
	Dashing becomes longer with increasing number of grid points in the first BZ ($2^M \times2^M$ points, with $M=1..5$). The solid line is the extrapolation for $M\to\infty$ and the dots are the exact result from an integration over the first BZ. 
Sufficiently far away from the phase transition, for a finite grid of roughly $32\times 32$ sites, $w_-$ can be reliably extracted. 
However, the result is not as reliable as the one using the method of Ref.~\cite{Fukui2005} [see main text, Fig.~3(f)]. 
	  \label{fig:nC_differencesums}
	}
\end{figure}
As seen in Fig.~\ref{fig:nC_differencesums}, this shortcoming can be partially remedied by a scaling with increasing resolution. 

A more robust method has been developed by Fukui and coworkers~\cite{Fukui2005}. In their approach, one defines link variables between points in $\bk$ space, 
\eq{
U_{\mu}(\bk)=\braket{n(\bk)|n(\bk+\hat{\mu})}/{\mathcal{N}_{\mu}(\bk)}\,, 
}
where $\ket{n(\bk)}$ is the eigenstate of the $n$'th band at quasimomentum $\bk$ and $\hat{\mu}$ the displacement vector by a discrete step in the BZ in direction of $\mu=x,y$. 
${\mathcal{N}_{\mu}(\bk)}=|\braket{n(\bk)|n(\bk+\hat{\mu})}|$ is a normalization. 
Using these link variables, one can define the lattice field strengths 
\eqa{
\label{eq:latticeFieldStrength}
\tilde{F}_{xy}(\bk)&=&\ln U_{x}(\bk) U_{y}(\bk+\hat{x}) U_{x}(\bk+\hat{y})^{-1}U_{y}(\bk)^{-1}\\
	&=&\ln U_{x}(\bk) U_{y}(\bk+\hat{x}) U_{-x}(\bk+\hat{x}+\hat{y}) U_{-y}(\bk+\hat{y})\nonumber
}
(where the logarithm is defined on the principal branch). 
The formulation in the second line corresponds to a Wilson loop of the $U$'s around a plaquette in $\bk$ space, which immediately shows its gauge invariance. 
From Eq.~\eqref{eq:latticeFieldStrength}, the Chern number can be approximated as 
\eq{
\tilde{w}=\frac{1}{2\pi i} \sum_{m} \tilde{F}_{xy}(\bk_{m})\,.
} 
An advantage of this discretization is that its result is always an integer \cite{Fukui2005}. 
Hence, it converges much faster towards the real Chern number than alternative methods. 
In fact, it can be shown that the number of $k$-space points needed in the first BZ to obtain the exact result is on the order of $2|w|$ \cite{Fukui2005}. 

Now, we want to show how the lattice field strengths \eqref{eq:latticeFieldStrength} can be computed using only the Hamiltonian vector $\hat{\bh}(\bk)$ as input. 
To do this, we rewrite Eq.~\eqref{eq:latticeFieldStrength} in terms of the projector onto the eigenstate $\ket{n(\bk)}$, $P_{\bk}^{(n)}=\ket{n(\bk)}\bra{n(\bk)}$, namely
\eq{
\label{eq:latticeFieldStrengthTraceOfProjectors}
\tilde{F}_{xy}(\bk)=\ln \left[ 
\frac{\Tr\left( P_{\bk} P_{\bk+\hat{x}} P_{\bk+\hat{x} +\hat{y}} P_{\bk+\hat{y}} \right) }
{
{\mathcal{N}_{x}(\bk)}  {\mathcal{N}_{y}(\bk+\hat{x})} {\mathcal{N}_{x}(\bk+\hat{y})}  {\mathcal{N}_{y}(\bk)} 
}
\right]\,, 
}
with ${\mathcal{N}_{\mu}(\bk)}=\sqrt{|\Tr ( P_{\bk} P_{\bk+\hat{\mu}})|} $. This formula is generally valid for non-Abelian Berry connections. 
It has the advantage that the projectors can be easily expressed in terms of the original Hamiltonian. 
For lattices with two sublattice states as they are considered here, the projectors are given by 
$P_{\bk}^{(\pm)}=\frac{1}{2}(\mathbb{I}\pm\hat{\bh}(\bk)\cdot\boldsymbol{\sigma})$. 
Therefore, knowledge of $\hat{\bh}(\bk)$ allows to immediately compute the Chern number. 
The result is presented in Fig.~3(f) of the main text (where we assumed that the additional information of the sign change of $\hh_z$ is known).\\

\subsection{Influence of a harmonic trap\label{sec:harmonic_trap}}
To date, most optical-lattice experiments are performed in the presence of a harmonic confinement (but see also the recent developments as in Ref.~\cite{Gaunt2013}). 
Such a harmonic confinement can reduce the contrast of the scheme, but the essentials of the method still work, as we discuss now. 

\subsubsection{Momentum-distribution in local density approximation}
In general, the interference pattern after a time of flight, if we neglect the Wannier envelope, is given by 
\eq{
\label{eq:nkoftgeneral}
n(\bk,t)=\sum_{\ell,\ell^\prime}\sum_{s,s^\prime} \ue^{-i\bk(\br_{\ell s}-\br_{\ell^\prime s^\prime})} \ue^{i(\nu_{\ell s}-\nu_{\ell^\prime s^\prime}) t } \braket{\hat{c}_{\ell s}^\dagger \hat{c}_{\ell^\prime s^\prime}}
}
In the absence of a trap, one has $\nu_{\ell s}=\nu_{s}$ and the correlators have to be evaluated in the state $\Pi_\bk c_{\bk-}^\dagger \ket{0}$, denoting a completely filled lower band (for simplicity, we assume a band insulator at vanishing temperature).  
Employing the Fourier transformation $\hat{c}_{\ell s}^\dagger=\frac{1}{\sqrt{N_u}}\sum_{\bk\in 1{\rm{st\, BZ}}} \left(\braket{\bk-|\ell s}\hat{c}_{\bk-}^\dagger+\braket{\bk+|\ell s}\hat{c}_{\bk+}^\dagger\right)$, where $N_u$ is the number of unit cells, one obtains 
\eq{
\braket{\hat{c}_{\ell s}^\dagger \hat{c}_{\ell^\prime s^\prime}}=\frac{1}{N_u}\sum_{\bk^\prime\in 1{\rm{st\, BZ}}} \braket{\ell^\prime s^\prime|\bk^\prime-}\braket{\bk^\prime-|\ell s}\,. 
}
Using 
\begin{subequations}
\eqa{
\braket{\ell A|\bk -}&=&\phantom{-}\frac{1}{\sqrt{N_u}}\ue^{-i\bk \cdot \br_{\ell A}}\sin\left(\frac{\vartheta_\bk}{2}\right)\,, \\
\braket{\ell B|\bk -}&=&-\frac{1}{\sqrt{N_u}}\ue^{-i\bk \cdot \br_{\ell B}}\ue^{i\varphi_\bk}\cos\left(\frac{\vartheta_\bk}{2}\right)\,,
}
\end{subequations}
one arrives at the formula (1) of the main text. 

In the presence of a trap, however, Eq.~(1) has to be modified. 
If we assume a shallow trap, so that we can employ a local density approximation, then its principal effect will be a local decrease of the Fermi energy as $\mu(\br_{\ell s})=\mu(r_{\ell s})=\mu_0-\frac{1}{2}m\omega_{\rm tr}^2 r_{\ell s}^2$. 
Here, $m$ is the atom mass and $\omega_{\rm tr}$ is the frequency of the trap, which we assumed cylindrically symmetric ($r_{\ell s}=|\br_{\ell s}|$). 
Further, $\mu_0$ is the Fermi energy in the centre of the trap. Then, the mode with energy $\epsilon_-(\bk)$ will be occupied within a radius of the trap given by $\mu(r_{\max}(\bk))=\epsilon_-(\bk)$, $r^2_{\max}(\bk)=2(\mu_0-\epsilon_-(\bk))/m\omega_{\rm tr}^2$. In other words, particles with lower energy occupy a larger extent of the trap. For maximal contrast, it is thus convenient to choose the Fermi energy in the trap centre just below the upper band, $\mu_0\approx \min_{\bk} \epsilon_+(\bk)$. 
We can incorporate the spatially varying Fermi level by modifying the correlators in Eq.~\eqref{eq:nkoftgeneral} to 
\eqa{
\braket{\hat{c}_{\ell s}^\dagger \hat{c}_{\ell^\prime s^\prime}}_{\rm tr}&=&\frac{1}{N_u}\sum_{\bk^\prime} \braket{\ell^\prime s^\prime|\bk^\prime-}\braket{\bk^\prime-|\ell s} \\
& &\quad\quad \theta\left(\mu(r_{\ell s})-\epsilon_-(\bk^\prime)\right)\theta\left(\mu(r_{\ell^\prime s^\prime})-\epsilon_-(\bk^\prime)\right)\,, \nonumber
}
where $\theta\left(\mu(r_{\ell s})-\epsilon_-(\bk^\prime)\right)$ is the Heavyside function. 
Additionally, the local energy is now modified to $\nu_{\ell s}=\nu_s + \mu(r_{\ell s})$. 

For a shallow trap, it is justified to approximate $\mu(r_{\ell s})\approx \mu(r_{\ell})$. We can then carry out the sums over the unit cells $\ell$ separately, and we obtain 
\begin{align}
\label{eq:nkwithtrap}
&n(\bk,t)_{\rm tr}=\sum_{\bk^\prime} \left|I(\bk,\bk^\prime)\right|^2 \\
&\times \left[1-\sin(\vartheta_{\bk^\prime}) \cos((\bk-\bk^\prime)(\br_{A}-\br_{B}) + \varphi_{\bk^\prime}+\omega(t-t_m))\right] \nonumber \,. 
\end{align} 
Here, we defined $\br_{s}$ as the position of the basis atom $s$ within a unit cell. 
As we will see, $\left|I(\bk,\bk^\prime)\right|^2$ is strongly peaked at $\bk=\bk^\prime$, so that we obtain the original expression without trap, modified by a time-dependent factor $\left|I(\bk)\right|^2$. 
To see this, we have to evaluate $I(\bk,\bk^\prime)$, which is given by 
\begin{align}
&I(\bk,\bk^\prime)=\frac{1}{N_u}\sum_\ell \ue^{-i(\bk-\bk^\prime)\br_\ell} \ue^{-i\frac{1}{2}m\omega_{\rm tr}^2 r_\ell^2 } \theta\left(\mu(r_{\ell})-\epsilon_-(\bk^\prime)\right) \nonumber\\ 
&\approx\frac{1}{\pi R_{\max}^2}\int \ud^2 r \ue^{-i(\bk-\bk^\prime)\br} \ue^{-i\frac{1}{2}m\omega_{\rm tr}^2 r^2 } \theta\left(\mu(r)-\epsilon_-(\bk^\prime)\right)  \,.
\end{align}
In the second line, we assured correct normalisation by defining the radius of the entire cloud, $R_{\max}^2\equiv \max_\bk 2(\mu_0-\epsilon_-(\bk))/m\omega_{\rm tr}^2 =  2(\mu_0-\epsilon_{\min})/m\omega_{\rm tr}^2$, where $\epsilon_{\min}=\min_\bk \epsilon_-(\bk)$. 
Using the series expansion of the second exponential, we can evaluate the integral to 
\eq{
\label{eq:Ikkp}
I(\bk,\bk^\prime)=\frac{i}{(\mu_0-\epsilon_{\min})t} \sum_{n=1}^\infty\frac{(i(\epsilon_-(\bk^\prime)-\mu_0)t)^n}{n!}C_n(\bk,\bk^\prime) \,.
}
Here, $C_n(\bk,\bk^\prime)= {_1F_2}\left(n;1,n+1;-\xi \right)$, where $_{p_1}F_{p_2}$ denotes the generalized hypergeometric function, and $\xi\equiv\frac{1}{4}(qa)^2\frac{\mu_0-\epsilon_-(\bk^\prime)}{E_{\rm{osc}}}$, with $q\equiv|\bk-\bk^\prime|$,  $a$ the lattice spacing, and $E_{\rm{osc}}=\frac{1}{2}m\omega_{\rm tr}^2 a^2$ an intrinsic energy scale of the trap. 
Equation \eqref{eq:nkwithtrap} together with \eqref{eq:Ikkp} is the central result of this section. We will now proceed to providing some further analytical insight by invoking some weak approximations to $I(\bk,\bk^\prime)$. Afterwards, we will repeat some of the numerical experiments of the main text, demonstrating that the proposed measurement scheme works also in the presence of a trap. 

\subsubsection{Zero-momentum-transfer and short-time approximations\label{sec:approximationsIkkprime}}

The mixture of different $\bk$-modes by $I(\bk,\bk^\prime)$ is typical for experiments carried out in traps. For shallow traps, however, such momentum transfers can typically be neglected. In Eq.~\eqref{eq:Ikkp}, this becomes manifest through the small denominator of $\xi$ (the generalised hypergeometric function decreases fast as a function of its argument). Thus, the integral $I(\bk,\bk^\prime)$ is strongly peaked at $\bk=\bk^\prime$ and it is justified to set $I(\bk,\bk^\prime)=I(\bk)\,\delta_{\bk,\bk^\prime}$. The experimental signature will therefore be
\begin{equation}
n(\bk,t)_{\rm tr}=\left|I(\bk)\right|^2 n(\bk,t) \,,
\end{equation} 
with $n(\bk,t)$ given by Eq.~(1) of the main text, and 
\eq{
\label{eq:Ikkprimeq0}
|I(\bk)|^2=  \left(\frac{\mu_0-\epsilon_-(\bk)}{\mu_0-\epsilon_{\min}}\right)^2 {\rm sinc}^2 \left(\frac{(\mu_0-\epsilon_-(\bk))t}{2}\right)\,.
}
Thus, the momentum distribution without trap obtains a multiplicative correction, with an intuitive interpretation: 
The first term in $|I(\bk)|^2$ is simply the reduction in contrast due to the reduced density at the borders of the trap (high-energy modes are only occupied in the centre of the trap where the Fermi energy is sufficiently high). 
The second term describes a dephasing of different regions of the trap, because during the time evolution after the quench the phase at each site oscillates with the site offset $\nu_s$ plus a locally varying contribution from the local chemical potential. 

Another useful limit of $I(\bk,\bk^\prime)$ that one can express in a simple analytical formula is the limit of short times, $t (\mu_0-\epsilon_-(\bk))\ll1$. Keeping terms up to $\mathcal{O}(t^2)$, we obtain 
\begin{align}
\label{eq:Ikkprimesmallt}
&|I(\bk,\bk^\prime)|^2=\left(\frac{\mu_0-\epsilon_{\bk^\prime}}{\mu_0-\epsilon_{\min}}\right)^2 \left\{
\frac 1 \xi \mathcal{J}_1(2\sqrt{\xi})^2 \right.\\
&\qquad\qquad- \left(\frac{(\mu_0-\epsilon_{\bk^\prime})t}{\xi}\right)^2 
\left[
\mathcal{J}_2(2\sqrt{\xi})^2-\sqrt{\xi}\mathcal{J}_3(2\sqrt{\xi}) \right. \nonumber\\
&\qquad\quad\qquad\qquad\qquad\qquad\left.\left.+(2-\xi)\mathcal{J}_1(2\sqrt{\xi})\mathcal{J}_3(2\sqrt{\xi})
\right]
\right\}\,, \nonumber
\end{align}
where $\mathcal{J}_\gamma(X)$ denote Bessel functions. 
Again, the same two effects as above are found, the overall loss of contrast due to the overall prefactor in front of the curly brackets and a gradual decrease in contrast due to the dephasing between different trap regions.  

In the numerics presented in the following section, we use realistic parameters, appearing in state-of-the-art experiments \cite{Aidelsburger2013,Struck2013}, $J/\hbar=2\pi\cdot 0.26\,\rm{kHz}$, $\omega=2\pi\cdot 10\,\rm{kHz}$, $\omega_{\rm tr}=2\pi\cdot 50\,\rm{Hz}$, $a=380\,\rm{nm}$. 
For these values, we have $E_{\rm osc}\approx 2\pi\cdot 1.6\,{\rm Hz} \ll (\mu_0-\epsilon_-(\bk))=\mathcal{O}(J)$ 
and, during one period, $t (\mu_0-\epsilon_-(\bk))\leq  \mathcal{O}(\frac{2\pi}{\omega}J) \ll 1$. 
Therefore, either approximation, Eq.~\eqref{eq:Ikkprimeq0} or Eq.~\eqref{eq:Ikkprimesmallt}, is well justified in a typical experimental setting with such parameters, especially the short-time approximation. 

\subsubsection{The $\pi$-flux Harper model in a trap}

Using the above parameter values, we recalculated Fig.~2 of the main text for the case of a trap, with the results shown in Fig.~\ref{fig:nkoftimeHarperTimeEvolutionTrap}. 
As can be seen in the upper row of Fig.~\ref{fig:nkoftimeHarperTimeEvolutionTrap}, the signal is deteriorated due to the population decrease close to the edges of the sample, especially for the modes around $k_x=\pi/2$, which have higher energy. 
However, as the bottom row demonstrates, the winding of $\hat{\bh}(\bk)$ can still be extracted reliably, at least far away from the topological transition \cite{footnoteExtractionOfh1}. We now turn to a more detailed analysis of the topological transition. 

\begin{figure}
	\centering
\includegraphics[width=0.5\textwidth]{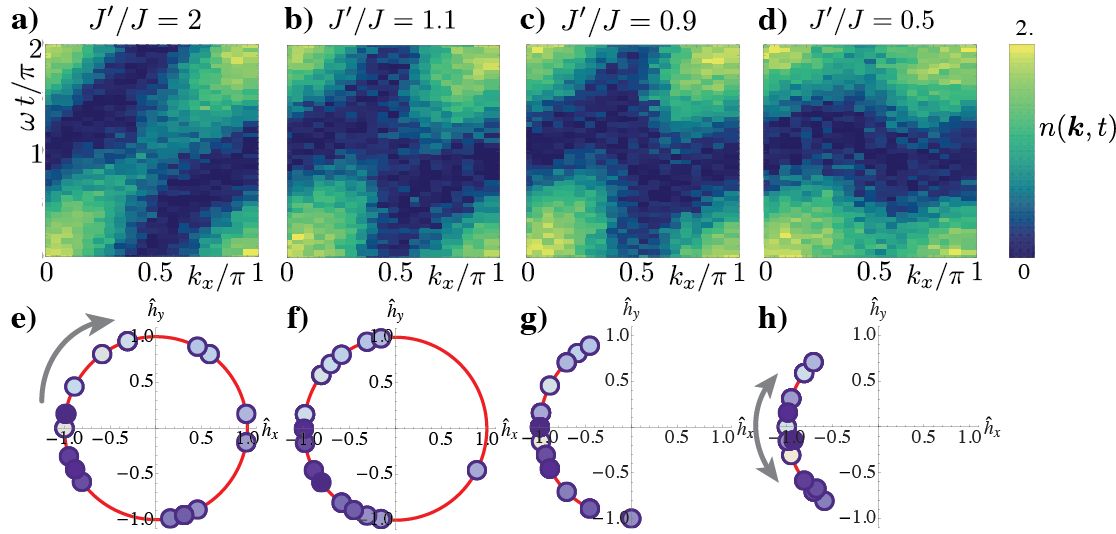}	
\caption{
As Fig.~2 of the main text, but in the presence of a harmonic trapping potential. 
(a-d) Time-resolved TOF images for the $\pi$-flux Harper model ($\Delta=0$ and $k_y^0=\pi/2$). 
We assumed the experimental parameters given at the end of Sec.~\ref{sec:approximationsIkkprime}, and took a realistic lattice size of $40\times 40$ occupied sites. 
Again, we added normal-distributed noise with a standard deviation of 0.1 of the average ideal signal.  
In a trap, the signal quality decreases due to the decrease of the number of particles towards the edge of the sample. 
However, despite the smaller signal, the presence (closed circle in e,f) or absence (open circle in g,h) of edge states is still visible. 
 \label{fig:nkoftimeHarperTimeEvolutionTrap}
	}
\end{figure}

In Fig.~\ref{fig:phaseTransitionHarperModel_trap}, we repeat the analysis of Fig.~\ref{fig:phaseTransitionHarperModel} for the topological transition as a function of $J^\prime/J$ \cite{footnoteExtractionOfh2}.
The blue bullets are data computed for the trap with the above parameters, under the addition of normal-distributed noise with a standard deviation of 0.1 of the average ideal signal. The extracted curve lies close to the dashed line, which represents the clean data in the presence of a trap, i.e., without the added noise. 
The trap also slightly shifts the topological transition from the trap-free case (solid line), but the general behaviour is clearly preserved. 
The overall behavior is similar to what happens, e.g., to the Mott-insulator--Superfluid transition, which also becomes less sharp in the presence of a trap. 
These results demonstrate that, even though a trap diminishes the quality of the results, the proposed scheme allows a reliable tomography of the Hamiltonian $\hat{\bh}(\bk)$. 
Let us also stress that the problems associated to a trap can be avoided altogether in a new generation of optical-lattice experiments, which work in box potentials without harmonic confinement Ref.~\cite{Gaunt2013}.

\begin{figure}
	\centering
	\includegraphics[width=0.29\textwidth]{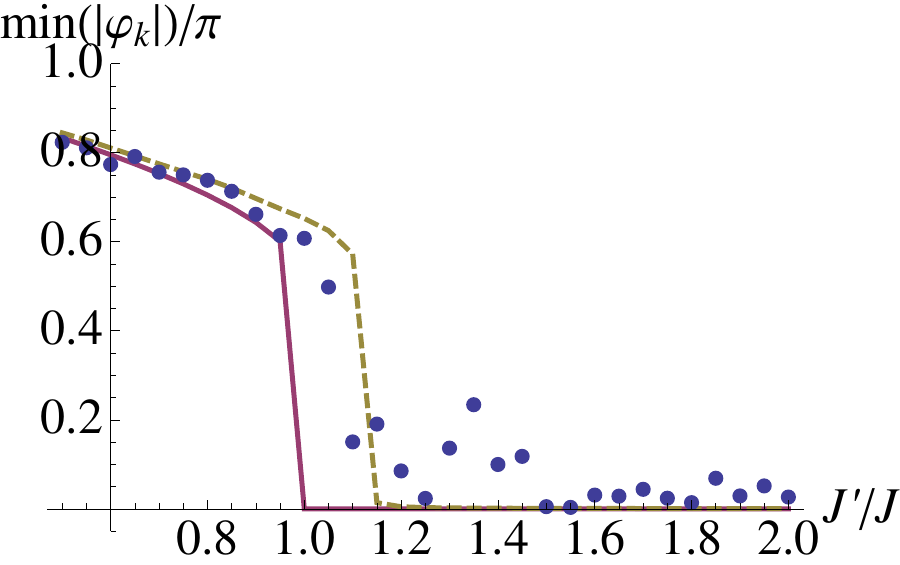}	
\caption{
	  {Topological phase transitions in the  $\pi$-flux Harper model in the presence of a trap.} 
		Edge modes at $k_y^0=\pi/2$ are supported if $\hat{\vect{h}}_{k_y^0}(k_x)$ encloses a circle around the origin, quantified by a vanishing $\min|\varphi_{k_x,k_y^0}|$. 
		The dots simulate a noisy experiment in the presence of a trap, with $40\times 40$ lattice sites, 41 time steps, and an added noise of 0.1 of the mean of the ideal case. 
		The appearance of zero-energy edge modes for $J^\prime/J>1$ is less clear than in the trap-free case (Fig.~\ref{fig:phaseTransitionHarperModel}a), but can still be distinguished.
		The dashed line represents the idealised prediction for an experiment in a trap but without any noise. It is slightly shifted from the ideal, trap-free case (solid line).  
	  \label{fig:phaseTransitionHarperModel_trap}}
\end{figure}


\begin{thebibliography}{44}
\bibitem {footnoteUnipartite}%
  \BibitemOpen
  Our measurement scheme itself is robust also when approaching the limit of an enhanced translational symmetry where both sublattice states become equivalent. 
  However, in this trivial limit the band gap closes, impeding the preparation of the band insulator
  \BibitemShut {NoStop}%
\bibitem {footnoteExtractionOfh1}%
  \BibitemOpen
  To obtain the form of $\hat{\bh}$ in these plots, we simply normalise the oscillation of $n(\bk,t)_{\rm tr}$ by its mean value over one time period, and then proceed similar as before, i.e., we identify $2\sin(\vartheta_\bk)$ with the amplitude of the normalised time-periodic oscillation and $\varphi_\bk$ with the shift of the first maximum
  \BibitemShut {NoStop}%
\bibitem {footnoteExtractionOfh2}%
  \BibitemOpen
To obtain the form of $\hat{\bh}$ for these data, we fit a cosine plus constant to $n(\bk,t)_{\rm tr}$, and normalise the results again by the mean value over one period
\BibitemShut {NoStop}%
\end{thebibliography}
\end{document}